\documentclass[aps,prb,showpacs,preprintnumbers,amsmath,amssymb,twocolumn,letterpaper]{revtex4}
\usepackage{amssymb}
\usepackage{ifpdf}
\usepackage[utf8]{inputenc}
\usepackage{graphicx}
\usepackage{hyperref}
\usepackage{bm}
\begin{document}
\title{Single magnetic chirality in the magneto-electric NdFe$_3$($^{11}$BO$_3$)$_4$}

\author{M. Janoschek}

\affiliation{Laboratory for Neutron Scattering, Paul Scherrer Institut \& ETH Zurich, CH-5232, Villigen, PSI, Switzerland}
\affiliation{Physik Department E21, Technische Universit\"at M\"unchen, D-85748 Garching, Germany}
\email{mjanoschek@physics.ucsd.edu}
\altaffiliation[Present address]{Department of Physics, University of California, San Diego, La Jolla, CA 92093-0354, USA}

\author{P. Fischer}

\affiliation{Laboratory for Neutron Scattering, Paul Scherrer Institut \& ETH Zurich, CH-5232, Villigen, PSI, Switzerland}

\author{J. Schefer}

\affiliation{Laboratory for Neutron Scattering, Paul Scherrer Institut \& ETH Zurich, CH-5232, Villigen, PSI, Switzerland}

\author{B. Roessli}

\affiliation{Laboratory for Neutron Scattering, Paul Scherrer Institut \& ETH Zurich, CH-5232, Villigen, PSI, Switzerland}

\author{V. Pomjakushin}

\affiliation{Laboratory for Neutron Scattering, Paul Scherrer Institut \& ETH Zurich, CH-5232, Villigen, PSI, Switzerland}

\author{M. Meven}

\affiliation{Forschungsneutronenquelle Heinz Maier-Leibnitz (FRM II), Technische Universit\"at M\"unchen, D-85748
Garching, Germany}

\author{V. Petricek}

\affiliation{Institute of Physics, ASCR v.v.i, Na Slovance 2, 182 21 Praha 8, Czech Republic}

\author{G. Petrakovskii}

\affiliation{Institute of Physics SB RAS, Krasnoyarsk 660036, Russia}

\author{L. Bezmaternikh}

\affiliation{Institute of Physics SB RAS, Krasnoyarsk 660036, Russia}

\date{\today}

\begin{abstract}
We have performed an extensive study of single-crystals of the magneto-electric NdFe$_3$($^{11}$BO$_3$)$_4$ by means of a combination of single-crystal neutron diffraction and spherical neutron polarimetry. Our investigation did not detect significant deviations at low temperatures from space group \textit{R32} concerning the chemical structure. With respect to magnetic ordering our combined results demonstrate that in the commensurate magnetic phase below T$_N$~$\approx$~30~K all three magnetic Fe moments and the magnetic Nd moment are aligned ferromagnetically in the basal hexagonal plane but align antiferromagnetically between adjacent planes. The phase transition to the low-temperature incommensurate magnetic structure observed at T$_{IC}$~$\approx$~13.5~K appears to be continuous. By means of polarized neutron studies it could be shown that in the incommensurate magnetic phase the magnetic structure of NdFe$_3$($^{11}$BO$_3$)$_4$ is transformed into a long-period antiferromagnetic helix with single chirality. Close to the commensurate-incommensurate phase transition third-order harmonics were observed which in addition indicate the formation of magnetic solitons.
\end{abstract}


\vskip2pc

\maketitle

\section{Introduction}

\begin{figure}[th!]
\includegraphics[width=.49\textwidth,clip=]{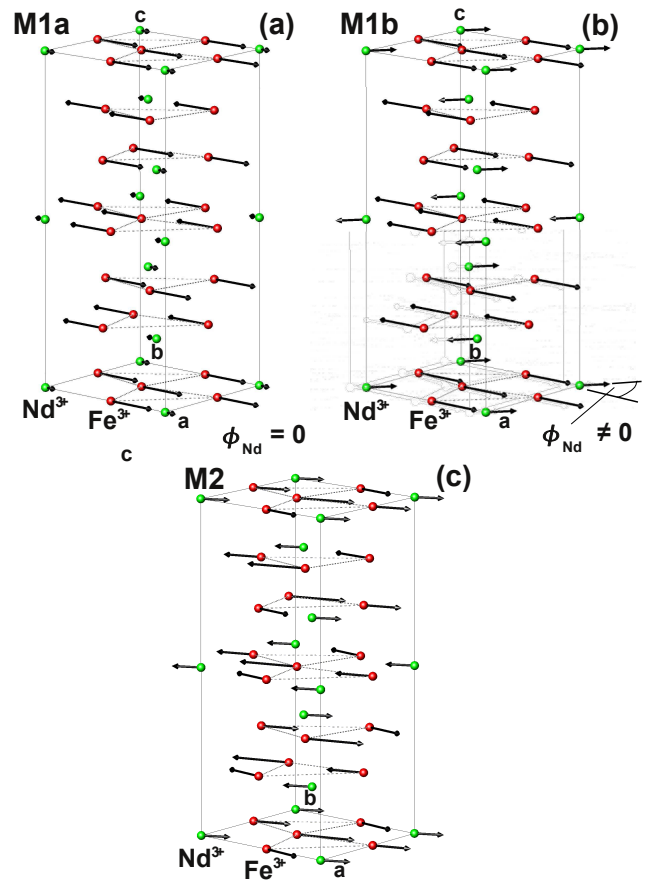}
\caption{Possible models for the magnetic structure of NdFe$_3$($^{11}$BO$_3$)$_4$ are plotted with the program ATOMS \cite{Dowty:06} for 20~K. We note that the magnetic unit cell is doubled along the crystallographic $c$-axis with respect to the chemical unit cell. Magnetic model (M1) features three magnetic Fe moments of equal magnitude. Panels (a) and (b) show two specific cases of (M1) where the angle $\phi_{\rm{Nd}}$ between the magnetic Fe and Nd moments is zero (M1a) or non-zero (M1b), respectively. The second magnetic model (M2) shown in (c) corresponds to  magnetic moments on the Fe sublattices that have slightly different sizes and orientations. Both models have been previously found to explain neutron powder diffraction data on NdFe$_3$($^{11}$BO$_3$)$_4$ by Fischer et al.\cite{fischer:06} equally well. Here we demonstrate that only the magnetic model (M1a) describes all present data correctly.}\label{Fig:old_mag_struc}
\end{figure}

Over the last decades various interesting effects related to long-range spiral-like forms of magnetic order have been revealed in multiple fields of condensed matter physics. Recent examples include multiferroic compounds such as TbMnO$_3$ \cite{kenzelmann:05,Kimura:07,Cheong:07}, magnetic surfaces ~\cite{Bogdanov:01,Bode:07} that are interesting for applications in spintronics, e.g. the construction of a spin field effect transistor\cite{Heide:06}, magnetic insulators such as Ba$_2$CuGe$_2$O$_7$ \cite{zheludev:96} or itinerant magnets such as MnSi\cite{Ishida:85}. Further, helimagnetism is also extensively discussed theoretically, here examples encompass the proposal of a chiral universality class \cite{Kawamura:98}, non-centrosymmetric superconductors \cite{Kaur:05} and new helical Goldstone modes, so-called helimagnons\cite{belitz:06,maleyev:06}. The latter have recently also been observed experimentally\cite{janoschek:09}.\\
Spiral magnetic order that in addition exhibits chirality, i.e. it breaks the spatial inversion symmetry, may exist in a right- and left-handed version that are interrelated via the inversion operation. As long as the underlying chemical structure is centrosymmetric both versions are energetically degenerate and right- and left-handed domains should be observed in equal fractions\cite{Brown:05, Bak:Jul80}. However, in case the chemical structure is non-centrosymmetric itself, the magnetic structure is expected to show a single handedness. Since 65 of the known 230 crystallographic space groups are non-centrosymmetric (Sohncke groups)\cite{flack:03} one would expect a great number of chiral magnetic compounds with a single chirality. Yet, only a few such compounds have been observed so far, with the most prominent example being MnSi. MnSi exhibits many interesting physical properties such as an extended non-Fermi liquid phase \cite{pfleiderer:04} probably associated with a new kind of metallic state \cite{pfleiderer:07,pfleiderer:07a}, a magnetic Skyrmion lattice \cite{muehlbauer:09} and the observation of helimagnons\cite{janoschek:09}, that are all closely related to its magnetic chirality\cite{Pfleiderer:09}. More recently monochirality was also observed in the non-centrosymmetric compound Ba$_3$NbFe$_3$Si$_2$O$_{14}$ that exhibits the interesting coexistence of two forms of chiral magnetism, within triangles in the basal plane and and along the $c$-axis as a magnetic helix\cite{Marty:08}. A further example is UPtGe that, however, displays a cycloid magnetic structure without true chirality, but with a single turning sense of the cycloid, that was also explained in terms of its non-centrosymmetric crystal structure\cite{mannix:00}.\\
In this work we demonstrate that NdFe$_3$($^{11}$BO$_3$)$_4$ is a new compound that displays a chiral magnetic structure with a single chirality. Below a magnetic commensurate (C) to incommensurate (IC) phase transition at $T_{IC}\approx$13.5~K the collinear antiferromagnetic structure of NdFe$_3$($^{11}$BO$_3$)$_4$ transforms into an antiferromagnetic helix that exists with a single chiral domain. Furthermore, the observation of third-order harmonics below $T_{IC}$ suggest that the emergence of the helical magnetic structure is accompanied by the formation of a magnetic soliton lattice below $T_{IC}$.\\
NdFe$_3$($^{11}$BO$_3$)$_4$ belongs to the family of borates RM$_3$(BO$_3$)$_4$ (R = Y,La-Lu, M = Al, Ga, Cr, Fe, Sc). These borates are interesting in their own right and have previously been studied mainly due to their special optical properties. Rare-earth ions, in general, and Nd$^{3+}$, in particular, have excellent characteristics to generate infrared laser action and to serve in nonlinear optics\cite{Jaque:01,Huang:02,Chen:01,chukalina:04}. The subfamily of ferroborates (M=Fe) is equally interesting with respect to their magnetic properties due to competing magnetic sublattices. Here GdFe$_3$(BO$_3$)$_4$ and NdFe$_3$(BO$_3$)$_4$ are especially fascinating as both materials show a large magneto-electric effect \cite{zvezdin:05,zvezdin:06}. \\
The ferroborates RFe$_3$(BO$_3$)$_4$ crystallize in the trigonal space group \textit{R32} (group no.~155), that is they belong to the structural type of the mineral huntite CaMg$_3$(BO$_3$)$_4$ \cite{campa:97}. Note that this structure is missing a center of inversion. In our previous work \cite{fischer:06} we performed an unpolarized neutron diffraction study with both powder and single crystal samples in order to  investigate the magnetic structure of NdFe$_3$($^{11}$BO$_3$)$_4$. We demonstrated that NdFe$_3$($^{11}$BO$_3$)$_4$ exhibits long-range antiferromagnetic order with the magnetic propagation vector $\bm{k}_{hex} = [0,0,3/2]$ below $T_N\approx$~30~K. However, our combined magnetic representational and Rietveld analysis\cite{rietveld:69} yields different magnetic structures that explain the data equally well (s. Fig. \ref{Fig:old_mag_struc}). All models have in common that the magnetic moments of all three Fe sublattices and the Nd sublattice are parallel to the hexagonal basal plane and are coupled antiferromagetically in adjacent planes. This is also in agreement with easy-plane type magnetic anisotropy that was observed in several studies of the magnetic susceptibility \cite{fischer:06, campa:97}. For a more detailed description we refer to Refs.~\onlinecite{fischer:06, janoschek:08}. In addition the study revealed that below approximately T~=~19~K the magnetic structure becomes incommensurate with the magnetic propagation vector $\bm{k}_{hex,i} = [0,0,3/2+\varepsilon]$. However, since the study was performed with thermal neutrons only, the experimental resolution did not allow for an exact study of the propagation vector as a function of temperature or the nature of the IC magnetic structure.
\\ 
The purpose of this study is mainly to identify the correct magnetic structure in both the C and IC magnetic phases. For this task we performed a detailed neutron diffraction study with both unpolarized neutrons and full spherical neutron polarimetry applied. Further high resolution diffraction was carried out to study the temperature dependence of the C-IC phase transition with better resolution. The unpolarized neutron single crystal diffraction data were additionally used to verify the low-temperature chemical structure of NdFe$_3$($^{11}$BO$_3$)$_4$, mostly as previous results indicated that the overall chemical symmetry might only be \textit{R3}~\cite{fischer:06}.\\
Finally all our magnetic single neutron diffraction results were evaluated by a comparative analysis with both the standard FullProf package\cite{fullprof} and Jana2006\cite{jana2006}. The latter was only recently extended for the purpose of analysis of magnetic neutron scattering data and emphasizes the use of magnetic symmetries. 

\section{Experimental techniques}

In this study two different samples of NdFe$_3$($^{11}$BO$_3$)$_4$ were studied. A large single crystal of approximate dimensions 8 x 8 x 8 mm$^3$  that has been already used in our previous work\cite{fischer:06} (Sample1) was reinvestigated. Further, a new smaller sample of the size 3 x 5 x 4mm$^3$ was prepared for additional experiments (Sample2). The process that was used for the preparation of the studied crystals is already described in detail in our previous publication\cite{fischer:06}.\\
In order to reduce strong extinction effects, diffraction measurements were carried out on the smaller Sample2 of NdFe$_3$($^{11}$BO$_3$)$_4$, in a four-circle setup on the thermal neutron diffractometer TriCS \cite{schefer:00} at the continuous Swiss spallation neutron source SINQ\cite{fischer:97} and on the HEiDi hot neutron diffractometer\cite{meven:05}, situated at the FRM II reactor at Munich. The corresponding neutron\cite{fischer:97} wavelengths were 1.18~\AA~ and 0.55~\AA, respectively. The data evaluations were performed by means of current versions of FullProf \cite{fullprof} and Jana2006 \cite{jana2006}. For the refinements the low-temperature lattice parameters a~=~9.594~\AA~ and c~=~7.603~\AA~ of Ref.~\onlinecite{fischer:06} were employed. Absorption corrections were neglected. Nuclear and magnetic extinction may be quite different, cf. e.g. \cite{Baruchel:86, brown:92}. We will show that in case of NdFe$_3$($^{11}$BO$_3$)$_4$ such a difference is almost not significant for hot neutrons, but turned out to be important for thermal neutrons. In contrast to FullProf, Jana2006 permits different types of isotropic extinction corrections. However, the corresponding differences were not found to be essential in case of NdFe$_3$($^{11}$BO$_3$)$_4$ and thus type II extinction\cite{becker:74} was used. In order to limit the number of parameters, only isotropic temperature factors were refined.\\
The recent extension of the Jana2006 refinement program allows commensurate and incommensurate magnetic structures to be described by four-dimensional magnetic superspace groups analogous to occupationally modulated chemical structures\cite{Janssen:06}. A more detailed discussion of these aspects of the refinement is given in the appendix together with a comparison to the corresponding FullProf results.\\
Further, Sample1 was investigated in a high resolution diffraction experiment carried out on the triple-axis spectrometer TASP\cite{semadeni:01}, situated at the end position of a cold super mirror guide of SINQ. The spectrometer was operated in its elastic mode with fixed incident and final wave vector k$_f$ = 1.2~\AA$^{-1}$. Additionally 20' Soller collimators were installed in the incident beam, in front of the analyzer and the detector. The second order contamination was removed from the beam by means of a beryllium filter that was inserted between the sample and the analyzer. The use of a triple-axis spectrometer for diffraction experiments is justified by the excellent signal-to-noise ratio that is achieved by the use of an additional analyzer crystal. This experimental setup only allows access to Bragg reflections in a single scattering plane. For this measurement the single crystal was oriented with the reciprocal axis b$^\ast$(K) and c$^\ast$(L) within the scattering plane.\\
In order to perform full polarization analysis on NdFe$_3$($^{11}$BO$_3$)$_4$, the spherical neutron  polarimetry (SNP) option MuPAD \cite{janoschek:07} available at SINQ was mounted on TASP. The neutron beam was polarized and analyzed via two polarizing supermirror benders that were installed after the monochromator and in front of the analyzer, respectively. A final wave vector k$_f$~=~1.97~\AA$^{-1}$ was chosen to maximize both the intensity and the polarization of the neutron beam. No additional filter for second order suppression was used because the benders already act as such. The orientation of the crystal was identical to the unpolarized measurements.\\
Only the combination of unpolarized and polarized neutron diffraction measurements allowed us to find the model for magnetic structure that gives the best agreement with all data collected during the course of this investigation. However, for the sake of clarity we will describe the unpolarized and the SNP measurements in separate sections, referring to the corresponding other sections when necessary.

\section{Unpolarized single crystal neutron diffraction}

\subsection{Low-temperature chemical structure}

The low temperature chemical structure was verified by measuring extended sets of nuclear neutron intensities on HEiDi at 22 K ($\lambda$~=~0.55 \AA, 287 Bragg peaks; data set N1) and on  TriCS at 6 K ($\lambda$~=~1.18~\AA, 151 Bragg peaks; data set N2). For the fits of both data sets with the space group \textit{R32} we refined 11 parameters  with two isotropic temperature factors (one for the heavy atoms Nd and Fe and one for the light atoms) and one isotropic extinction parameter. The resulting structural parameters are summarized in Table~\ref{tab:structure} together with the agreement factors. Note, that for the data set (N1) we performed fits with both FullProf and Jana2006. This is discussed in more detail in the appendix.\\
Our findings confirm that also at low temperatures the chemical structure of NdFe$_3$($^{11}$BO$_3$)$_4$ is well described within the space group \textit{R32}. The positional parameters derived from the two data sets are in good agreement. They are close to the room-temperature values published in our previous work\cite{fischer:06}. In contrast to the TriCS measurement, but in agreement with previous powder investigations \cite{fischer:06}, the HEiDi single crystal refinement tended for the isotropic temperature factor of Nd to negative values. Compared to the hot neutron results, the larger temperature factor values obtained from the thermal neutron data are presumably due to the neglected larger absorption in the latter case. Refinement of occupation factors did not indicate essential deviations from stoichiometry.

% Table I. 
\begin{table}[!ht]
\caption{Low-temperature structural parameters of NdFe$_3$($^{11}$BO$_3$)$_4$, refined from the present single crystal neutron diffraction data. Results from the fits of data sets (N1) are shown in the first and second lines were performed with FullProf (N1F) and Jana2006 (N1J), respectively. The third line gives the results from data set (N2) performed with Fullprof. Standard uncertainties of the parameters are given within parentheses. The agreement factors for the corresponding fits are: (N1F)  R$_{n,F2}$ = 4.5 \%, R$_{n,F2w}$ = 5.3 \%, R$_{n,F}$ = 3.2 \%, goodness of fit $\chi^2$ = 6 (s. Ref.~\onlinecite{fullprof} for the meaning of agreement factors); (N1J) R(all) = 3.1 \%, R$_w$(all) = 6.5 \%, GOF(all) = 1.7, GOF(obs) = 1.8 (s. Ref.~\onlinecite{jana2006}); (N2) R$_{n,F2}$ = 8.0 \%, R$_{n,F2w}$ = 8.4 \%, R$_{n,F}$ = 5.6 \%, goodness of fit $\chi^2$ = 258.}\label{tab:structure}
\centering
{\footnotesize
\begin{tabular}{l|lll|l}
\hline
Atom	& x		& y		& z		& B (\AA$^2$)\\\hline\hline
Nd	& 0		& 0		& 0		& 0.00(4)    \\
	&		&		&		& 0.0(2)     \\
	&		&		&		& 1.03(7)    \\\hline	
Fe	& 0.5506(2)	& 0		& 0		& 0.00(4)    \\
        & 0.5506(2)     &               &               & 0.0(2)     \\
	& 0.5505(3)     &               &               & 1.03(7)    \\\hline
B1	& 0		& 0		& 0.5           & 0.21(3)    \\
	&		&		&		& 0.17(4)    \\\
	&		&		&		& 1.14(6)    \\\hline	
B2      & 0.4465(3)     & 0             & 0.5           & 0.21(3)    \\
        & 0.4466(3)     &               &               & 0.17(4)    \\\
        & 0.4468(5)     &               &		& 1.14(6)    \\\hline	
O1      & 0.8565(4)     & 0             & 0.5           & 0.21(3)    \\
        & 0.8567(5)     &               &               & 0.17(4)    \\\
        & 0.8586(8)     &               &		& 1.14(6)    \\\hline	
O2      & 0.5901(3)     & 0             & 0.5           & 0.21(3)    \\
        & 0.5901(4)     &               &               & 0.17(4)    \\\
        & 0.5903(6)     &               &		& 1.14(6)    \\\hline	
O3      & 0.4523(3)     & 0.1445(3)     & 0.5185(3)     & 0.21(3)    \\
        & 0.4521(3)     & 0.1449(3)     & 0.5180(4)     & 0.17(4)    \\\
	& 0.4533(5)     & 0.1448(5)     & 0.5182(4)     & 1.14(6)    \\\hline	
\end{tabular}}
\end{table}

\subsection{Magnetic commensurate-incommensurate transition}\label{highres}

\begin{figure}[bh]
\centering
\includegraphics[width=.42\textwidth,clip=]{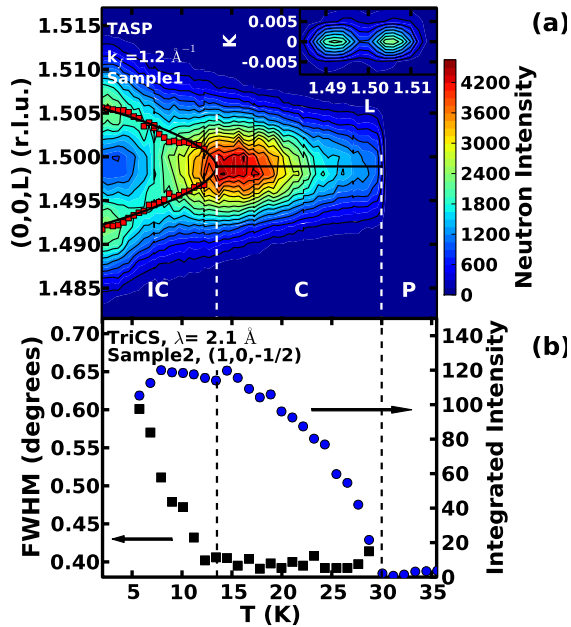}
\caption{(a) The figure shows a contour map of the neutron intensity around the reciprocal lattice position (0,0,$\frac{3}{2}$) of Sample1 in elastic $\bm{Q}$-scans along the reciprocal L-direction as a function of temperature determined on the triple-axis spectrometer TASP. The red squares are the peak positions of the incommensurate magnetic Bragg reflections as determined by fits of Gaussian peaks to the measured scans.  Here (P) denotes the paramagnetic phase, (C) the commensurate magnetic phase and (IC) the incommensurate magnetic phase, respectively. The inset shows a contour map of the neutron intensity around the position (0,0,$\frac{3}{2}$) in reciprocal space determined at T~=~1.6~K. (b) Temperature dependencies of the integrated neutron intensity and of the full width at half maximum (FWHM) of the magnetic Bragg peak (1,0,-1/2) of Sample2 are shown.}\label{fig:k_tempdep}
\end{figure}

\begin{figure}[bh!]
\includegraphics[width=.38\textwidth,clip=]{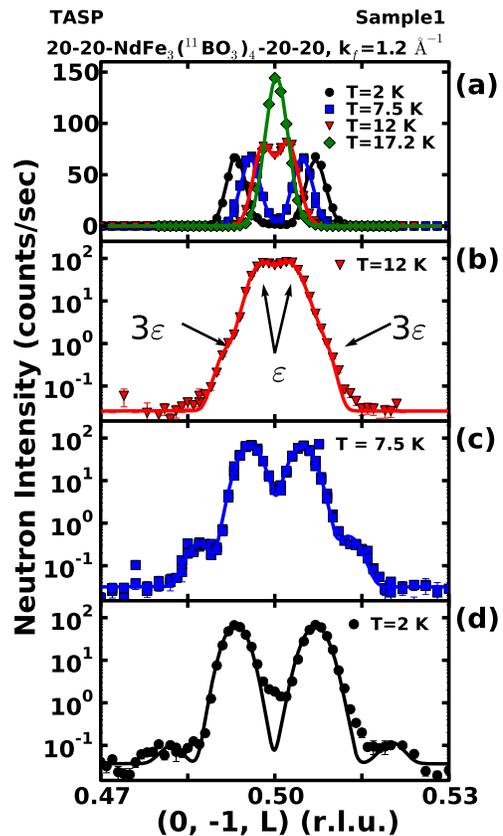}
\caption{(a) $\bm{Q}$-scans over the (0,-1,0.5) magnetic Bragg for different temperatures are shown. The scans were performed with 20' Soller collimators installed in the incident beam, in front of the analyzer and the detector. In the panels (b) to (d) the development of third order satellites is demonstrated. Their shift from the commensurate position is three times larger than for the first order peaks. The solid lines are fits to the data with multiple Gaussian profiles. }\label{Fig:nd_second_order_sats}
\end{figure}

By means of careful cold neutron scattering experiments on the triple-axis spectrometer TASP carried out on the bigger Sample1, we investigated the temperature dependence of the magnetic C-IC transition. Elastic $\bm{Q}$-scans along the reciprocal L-direction around the magnetic Bragg reflection (0,0,$\frac{3}{2}$) as a function of temperature were used to determine the transition temperature as illustrated in Fig.~\ref{fig:k_tempdep}(a). The propagation vector is commensurate down to the temperature T$_{IC}$~$\approx$~13.5~K below which the magnetic Bragg reflection splits into two incommensurate satellite peaks. Here we determined T$_{IC}$ as the temperature where the maximum intensity of the commensurate magnetic reflection is reached and then starts to decrease as the reflection splits up into the two incommensurate satellite peaks.
Below T$_{IC}$ the scans show that $k_z$ changes continuously at the C-IC phase transition in NdFe$_3$($^{11}$BO$_3$)$_4$ and can be well described via $k_z = \frac{3}{2} + \varepsilon$ where 
\begin{equation}
 \varepsilon = 1.6\cdot 10^{-3}\vert(T_{IC}-T)\vert^{0.58}.\label{eq:k_temp_dep}
\end{equation}
This is demonstrated by the black solid line in Fig.~\ref{fig:k_tempdep}(a). The splitting of the magnetic Bragg reflections is only observed along the $z$-direction down to T~=~1.6~K as shown by the intensity map around the (0,0,$\frac{3}{2}$) provided in the inset of  Fig.~\ref{fig:k_tempdep}(a). The splitting at T~=~1.6~K is $\varepsilon$ = 0.00667.\\
In addition to the principal magnetic satellites also reflections at higher order harmonics were observed as shown in Fig.~\ref{Fig:nd_second_order_sats}. Their shift with respect to $\bm{k}_{hex}$ was found to be 3$\varepsilon$~=~0.02. The integrated intensity of the third order harmonics is largest near the transition temperature T$_{IC}$ and decreases fast as a function of decreasing temperature. We note that the integrated intensity of the principle satellites remains approximately constant down the lowest temperature as would be expected for a magnetic soliton lattice~\cite{roessli:01}.\\
Investigations performed on TriCS with thermal neutrons on the smaller Sample2 of NdFe$_3$($^{11}$BO$_3$)$_4$ clearly show broadening of magnetic peaks at temperatures below 12.5~K, confirming the existence C-IC magnetic phase transition in NdFe$_3$($^{11}$BO$_3$)$_4$. This is illustrated in Fig.~\ref{fig:k_tempdep}(b). However, a splitting of the (-1,0,0.5) peak as observed before on the Sample1\cite{fischer:06}, could not be reproduced under similar conditions or on the HEiDi neutron diffractometer. This indicates a certain sample dependence of the k-vector magnitude in the incommensurate phase. At the lowest temperature $k_z$ may be approximated as 1.502 for Sample2.\\


\subsection{Magnetic structure}\label{unpol_magstruc}

\begin{figure}[bh]
\centering
\includegraphics[width=.38\textwidth,clip=]{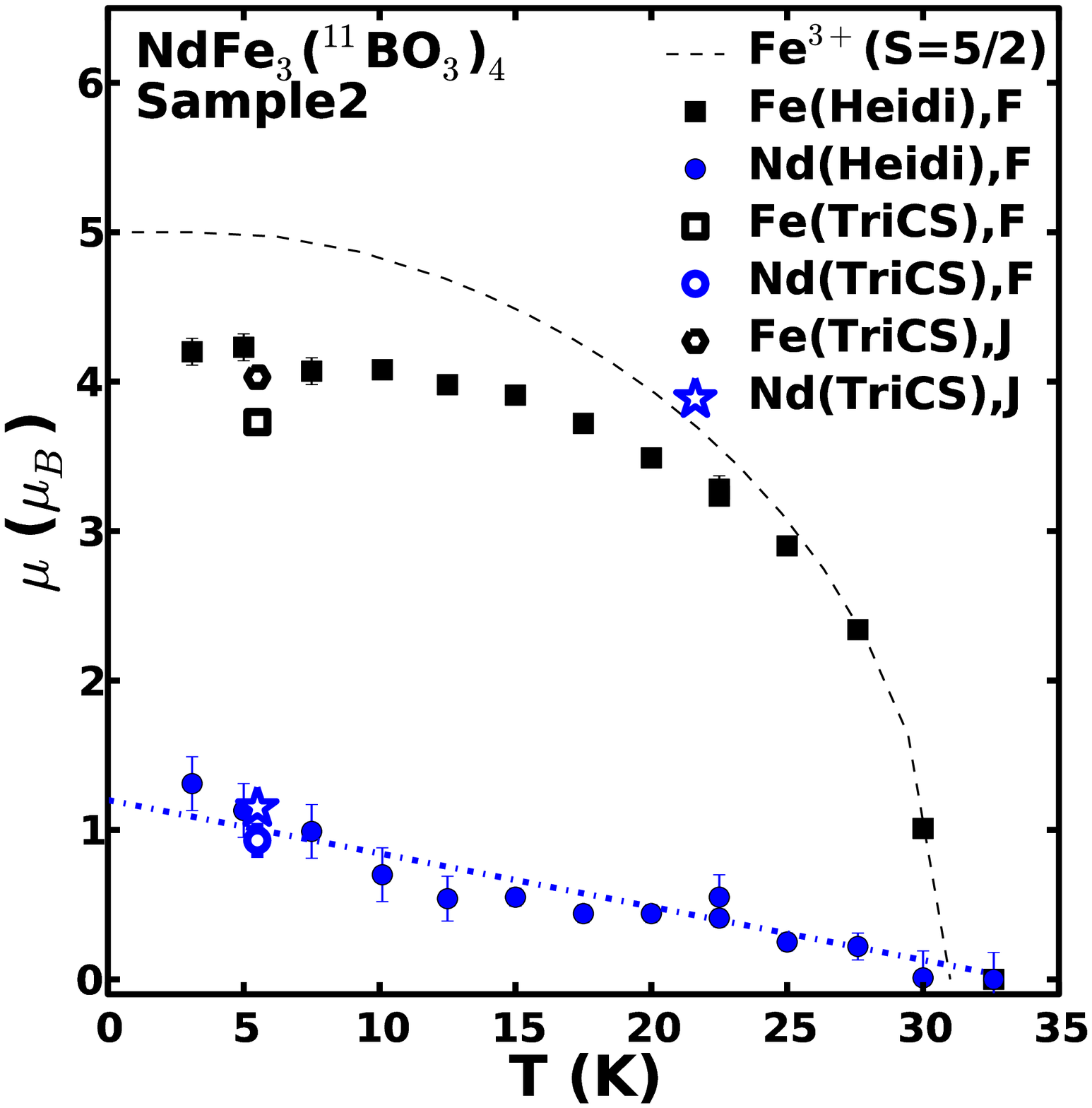}
\caption{Temperature dependencies of the ordered magnetic Fe and Nd moments in NdFe$_3$($^{11}$BO$_3$)$_4$  based on the evaluation of the present single crystal neutron diffraction data with FullProf (F). The values found via the alternative data refinement by means of Jana2006 (J) agree within the errorbars (cf. appendix) apart for the data measured with thermal neutrons on the instrument TriCS. The dashed line represents the one expected for S = 5/2 of Fe$^{3+}$ (cf. Ref.~\onlinecite{fischer:06}). The linear dashed-dotted line through the Nd moments is a guide to the eye. 
}\label{fig:magmom}
\end{figure}

Corresponding to the results in the preceding section, we assumed commensurate and incommensurate magnetic ordering for temperatures above and below T$_{IC}$, respectively, for the fits of the data discussed in the following. Moreover, we used the structural parameters from (N1) (HEiDi results).\\
The spherical neutron polarimetry experiments that will be discussed in section~\ref{SNP} showed that only the  model (M1) for the magnetic structure is able to explain the data correctly. Therefore, we will focus on the fits performed for this model here. As shown in Refs.~\onlinecite{fischer:06,janoschek:08} for (M1) the magnetic structure can be conveniently expressed as spirals of the form
\begin{equation}
  \begin{split}
  \bm{S}_j(\bm{t})=&S_{j}\Big[\bm{e}_x\cos(2\pi\bm{k\cdot t} +\phi_{j})+\\
  &(\bm{e}_x+2\bm{e}_y)/\sqrt{3}\sin(2\pi\bm{k\cdot t}+\phi_{j})\Big],
  \end{split}
  \label{eq:nd_fe_spiral} 
\end{equation}
that propagate along the hexagonal $c$-axis for all three magnetic Fe moments and the magnetic Nd moment ($j=Fe,Nd$) and also for both the (C) and (IC) magnetic phases. Here $\bm{e}_x$ and $\bm{e}_y$ are unit vectors of the hexagonal lattice and $\phi_j$ describes the polar takeoff angle from the hexagonal $a$ axis.\\
In the incommensurate case we superimposed $+$ and $-$ satellites in each magnetic peak, as here $\pm\bm{k}_{hex,i}$ are inequivalent. Further, we note, that for such spirals the magnetic neutron intensity is modulated with the factor $(1+\cos2\eta)$, where $\eta$ represents the angle of the magnetic scattering vector to the spiral axis\cite{lyons:62}. The direction of the three parallel magnetic Fe moments, that is described via the polar angle $\phi_{\rm{Fe}}$ in our model, cannot be determined from the present unpolarized neutron diffraction data. However, recent bulk magnetic measurements of Tristan et al. \cite{Tristan:07} on a single crystal of NdFe$_3$($^{11}$BO$_3$)$_4$ at low temperatures show easy magnetization along the $a$-axis and we therefore fixed the value of the polar angle to $\phi_{\rm{Fe}}$~=~0. In our previous work\cite{fischer:06} we found a small angle between magnetic moments of the Fe and Nd ions. Hence, we performed fits in two configurations, respectively, where we either kept the Nd magnetic moment parallel to the Fe moments (model (M1a)) or allowed for a refinable angle $\phi_{\rm{Nd}}$ between the Fe and Nd magnetic moments~(model M1(b)). Starting with a single magnetic domain and using the nuclear isotropic extinction parameters, thus 2 or 3 parameters were refined ((M1a) and (M1b), respectively). In principle up to six magnetic orientation domains would be possible according to the threefold- and twofold rotation axes of the paramagnetic space group \textit{R32} in the latter case: $u,v,w$; $-v,u-v,w$; $-u+v,-u,w$; $u-v,-v,-w$; $-u,-u+v,-w$; $v, u,-w$ in direct space notation. For the incommensurate case only the three threefold rotations around the crystallographic $c$-axis are present\cite{fischer:06}. However, the corresponding FullProf calculation did not yield improved fits. This can be understood by considering that the incommensurate magnetic spiral described by Eq.~\eqref{eq:nd_fe_spiral} breaks the translation symmetry of the underlying chemical lattice along the c direction. The broken translation symmetry implies that any rotation of the spiral around the $c$-axis can be compensated by a translation and the spin structure is conserved. The additional fits by means of Jana2006 gave identical results. In particular the superspace description for the magnetic structure in Jana2006 gives a natural explanation concerning the single populated magnetic orientation domain(s. appendix and Ref.~\onlinecite{petricek:09}).\\
In the commensurate case with $\bm{k}_{hex} = [0,0,3/2]$,  $+\bm{k}_{hex}$ is equivalent to $-\bm{k}_{hex}$ and this implies that the magnetic moments in adjacent planes are arranged antiparallel. Here the refinement according to model (M1a) yield three in good approximation equally distributed magnetic domains according to the threefold rotation axes.\\
At various temperatures sets of 50 magnetic and 5 nuclear Bragg peaks were measured on HEiDi. Corresponding characteristic refinement results obtained by both FullProf and Jana2006 are given in the appendix and are discussed subsequently.\\ 
For the commensurate magnetic phase of NdFe$_3$($^{11}$BO$_3$)$_4$ we may conclude $\phi_{\rm{Fe}}$~=~$\phi_{\rm{Nd}}$~=~0 according to model (M1a). We note our previously published powder neutron diffraction data\cite{fischer:06} may be equally well fitted with $\phi_{\rm{Fe}}$~=~$\phi_{\rm{Nd}}$~=~0 (cf. Ref.~\onlinecite{janoschek:08} additionally). Model (M1a) is furthermore supported by the spherical neutron polarimetry results that will be discussed in section ~\ref{SNP}.\\
With respect to incommensurate magnetic ordering in NdFe$_3$($^{11}$BO$_3$)$_4$, the thermal neutron data at 6 K indicate almost significance for the introduction of a non-zero angle $\phi_{\rm{Nd}}$(M1b). On the other hand, the difference between the two different HEiDi refinements is considerably smaller, due to less important extinction corrections at shorter neutron wavelength. Thus the simpler magnetic structure with $\phi_{\rm{Fe}}$~=~$\phi_{\rm{Nd}}$~=~0 (M1a) holds most probably also in the incommensurate phase of NdFe$_3$($^{11}$BO$_3$)$_4$ (s. appendix for details).\\
The resulting temperature dependencies of the ordered magnetic moments of the Fe$^{3+}$ and Nd$^{3+}$ ions in NdFe$_3$($^{11}$BO$_3$)$_4$ are shown in Fig.~\ref{fig:magmom}. The deviations between the magnetic moments derived from the 5.5 K TriCS data and those from HEiDi at 5 K are presumably due to the larger extinction corrections in case of thermal neutrons. The magnetic moments of both Fe and Nd seem to vary smoothly as a function of temperature, also at the C-IC magnetic phase transition at approximately T$_{IC}$~$\approx$~13.5~K. 

\section{Spherical neutron polarimetry}\label{SNP}

SNP measurements were performed above and below the C-IC transition, respectively. In the IC phase of NdFe$_3$($^{11}$BO$_3$)$_4$ an excellent $\bm{Q}$-resolution is necessary to observe the small splitting ($\varepsilon$ = 0.00667 (r.l.u.) $\equiv{}$ 0.0055\AA$^{-1}$) of the magnetic Bragg reflections. For the small final wave vectors k$_f$ that are required for such a high resolution setup, the polarizing benders used at TASP perform non-ideal with respect to transmission and polarizing efficiency. The best tradeoff between resolution and transmission/polarization is reached when TASP is operated with   k$_f$~=~1.97~\AA$^{-1}$. However, the relatively moderate $\bm{Q}$-resolution associated with k$_f$ = 1.97~\AA$^{-1}$ is not sufficient to perform separate SNP measurements on the magnetic satellites $\pm\bm{k}_{hex,i}$ and consequently a superposition of intensities from both satellites will be observed. 
Nevertheless important information could be extracted by performing polarized $\bm{Q}$-scans over the magnetic satellites.\\
We first discuss the results obtained by conventional SNP measurements in the commensurate phase of NdFe$_3$($^{11}$BO$_3$)$_4$ in section~\ref{SNP_C}, the findings for the incommensurate phase are described in section section~\ref{SNP_IC}.

\begin{table*}[th!]
\caption{Polarization matrices on all accessible magnetic Bragg reflections of NdFe$_3$($^{11}$BO$_3$)$_4$ are shown for T = 20~K. The column $\bm{P}_{0}$ and $\bm{P}'$ denote the direction of the initial and final polarization vector, respectively. The subscripts of $\bm{P}'$ indicate the polarization matrices that were measured (\textit{meas}) and calculated (\textit{calc}) from the two distinct magnetic models models (M1a) and (M2). The polarization tensor elements marked in bold demonstrate where model (M2) does not match the data.}
\label{tab:polartensors}
\centering       %p{4mm}p{4mm}p{5.5mm}
\vspace{2mm}
 {\scriptsize
 \tabcolsep1mm
 \begin{tabular}{rrr|r|ccc|ccc|ccc}\hline
  \multicolumn{3}{l|}{}&            &\multicolumn{3}{c|}{$\bm{P}_{meas}'$}&\multicolumn{3}{c|}{$\bm{P}_{calc}'(M1a)$} &\multicolumn{3}{c}{$\bm{P}_{calc}'(M2)$}\\
  H      &K     &L     &$\bm{P}_{0}$&    $x$     &     $y$      &      $z$     &      $x$     &     $y$      &      $z$       &    $x$     &     $y$      &      $z$  \\\hline\hline
%---------------------------------------------------------------------------------------------------------
  $+0.0$ &$+2.0$& $-0.5$&    $+x$    & $-0.861(3) $ & $+0.073(6) $ & $+0.042(6) $  & $-0.869$  &$+0.000$    &    $+0.000$   &  $-0.868$  &$+0.000$    & $+0.000$ \\
	 &      &       &    $+y$    & $-0.059(6) $ & $-0.659(4) $ & $-0.050(6) $  & $+0.000$  &$-0.756$    &  $+0.000$  &  $+0.002$  &$-0.752$    & $+0.000$ \\
         &      &       &    $+z$    & $+0.043(6) $ & $-0.050(6) $ & $+0.655(4) $  & $+0.000$  &$+0.000$    & $+0.756$  &  $+0.002$  &$+0.000$    & $+0.752$ \\

\cline{4-13}
%----------------------------------------------------------------------------------------------------------------------------------------------
         &      &       &    $-x$    & $+0.876(3) $ & $-0.026(6) $ & $-0.031(6) $  & $+0.869$  &$+0.000$    & $+0.000$   &  $+0.869$  &$+0.000$    & $-0.000$ \\
         &      &       &    $-y$    & $+0.034(6) $ & $+0.659(4) $ & $+0.059(6) $  & $+0.000$  &$+0.756$    & $+0.000$   &  $+0.002$  &$+0.752$    & $+0.000$  \\
         &      &       &    $-z$    & $-0.081(6) $ & $+0.043(6) $ & $-0.655(4) $  & $+0.000$  &$+0.000$    & $-0.756$   &  $+0.002$  & $+0.000$     & $-0.752$ \\

\hline
%----------------------------------------------------------------------------------------------------------------------------------------------
  $+0.0$ &$+0.0$& $-1.5$&    $+x$    & $-0.872(2) $ & $+0.057(4) $ & $+0.119(4) $  & $-0.869$  &$+0.000$    & $+0.000$   &  $-0.869$  &$+0.000$    & $+0.000$ \\
         &      &       &    $+y$    & $-0.010(4) $ & $+0.010(4) $ & $+0.019(4) $  & $-0.000$  &$+0.000$    & $-0.000$   &  $+0.000$  &$+0.000$    & $+0.000$ \\
         &      &       &    $+z$    & $-0.032(4) $ & $+0.020(4) $ & $-0.007(4) $  & $+0.000$  &$-0.000$    & $-0.000$   &  $+0.000$  &$+0.000$    & $-0.000$ \\

\cline{4-13}
%----------------------------------------------------------------------------------------------------------------------------------------------
         &      &       &    $-x$    & $+0.872(2) $ & $-0.058(3) $ & $-0.115(3) $  & $+0.869$  &$+0.000$    & $+0.000$   &  $+0.869$  &$+0.000$    & $+0.000$ \\
         &      &       &    $-y$    & $-0.043(4) $ & $-0.018(4) $ & $-0.017(4) $  & $-0.000$  &$-0.000$    & $+0.000$   &  $+0.000$  &$-0.000$    & $-0.000$ \\
         &      &       &    $-z$    & $-0.019(4) $ & $-0.023(4) $ & $+0.017(4) $  & $+0.000$  &$+0.000$    & $+0.000$   &  $+0.000$  &$-0.000$    & $+0.000$ \\

\hline
%----------------------------------------------------------------------------------------------------------------------------------------------
  $+0.0$ &$+4.0$& $+0.5$&    $+x$    & $-0.853(7) $ & $+0.06(1)  $ & $+0.01(1)  $  & $-0.869$  &$+0.000$    & $+0.000$   &  $-0.869$  &$+0.000$    & $+0.000$ \\
         &      &       &    $$+y$$    & $-0.05(1)  $ & $-0.837(8) $ & $-0.05(1)  $  & $+0.000$  &$-0.869$    & $+0.000$   &  $-0.004$  &$-0.866$    & $+0.000$ \\
         &      &       &    $+z$    & $+0.03(1)  $ & $-0.02(1)  $ & $+0.826(8) $  & $+0.000$  &$+0.000$    & $+0.869$   &  $-0.004$  &$+0.000$    & $+0.866$ \\

\cline{4-13}
%----------------------------------------------------------------------------------------------------------------------------------------------
         &      &       &    $-x$    & $+0.850(7) $ & $+0.02(1)  $ & $+0.03(1)  $  & $+0.869$  &$+0.000$    & $+0.000$   &  $+0.868$  &$+0.000$    & $+0.000$ \\
         &      &       &    $-y$    & $-0.01(1)  $ & $+0.807(8) $ & $+0.08(1)  $  & $+0.000$  &$+0.869$    & $-0.000$   &  $-0.004$  &$+0.866$    & $+0.000$ \\
         &      &       &    $-z$    & $-0.08(1)  $ & $+0.08(1)  $ & $-0.851(7) $  & $+0.000$  &$-0.000$    & $-0.869$   &  $-0.004$  &$-0.000$    & $-0.866$ \\

\hline
%----------------------------------------------------------------------------------------------------------------------------------------------
  $+0.0$ &$-2.0$& $-2.5$&    $+x$    & $-0.871(4) $ & $+0.068(7) $ & $+0.107(7) $  & $-0.869$  &$-0.000$    & $+0.000$   &  $-0.869$  &$-0.000$    & $+0.000$ \\
         &      &       &    $$+y$$    & $-0.092(7) $ & $-0.081(7) $ & $-0.012(7) $  & $+0.000$  &$-0.184$    & $-0.001$   &  $-0.003$  &$-0.170$    & $-0.000$ \\
         &      &       &    $+z$    & $-0.017(7) $ & $-0.006(7) $ & $+0.100(7) $  & $+0.000$  &$+0.001$    & $+0.184$   &  $-0.003$  &$-0.000$    & $+0.170$ \\

\cline{4-13}
%----------------------------------------------------------------------------------------------------------------------------------------------
         &      &       &    $-x$    & $+0.876(4) $ & $-0.045(7) $ & $-0.102(7) $  & $+0.869$  &$+0.000$    & $+0.000$   &  $+0.868$  &$-0.000$    & $-0.000$ \\
         &      &       &    $-y$    & $-0.005(7) $ & $+0.085(7) $ & $+0.030(7) $  & $+0.000$  &$+0.184$    & $-0.001$   &  $-0.003$  &$+0.170$    & $+0.000$ \\
         &      &       &    $-z$    & $-0.064(7) $ & $+0.030(7) $ & $-0.087(7) $  & $+0.000$  &$-0.001$    & $-0.184$   &  $-0.003$  &$+0.000$    & $-0.170$ \\

\hline
%----------------------------------------------------------------------------------------------------------------------------------------------
  $+0.0$ &$+1.0$& $-2.5$&    $+x$    & $-0.860(8) $ & $+0.07(2)  $ & $+0.11(2)  $  & $-0.869$  &$+0.000$    & $+0.000$   &  $-0.898$  &$+0.000$    & $-0.000$ \\
         &      &       &    $$+y$$    & $-0.04(2)  $ & $-0.08(2)  $ & $+0.01(2)  $  & $+0.000$  &$-0.055$    & $+0.000$   &  $\bm{-0.285}$  & $\bm{-0.649}$    & $+0.000$ \\
         &      &       &    $+z$    & $-0.04(2)  $ & $-0.01(2)  $ & $+0.11(2)  $  & $ 0.000$  &$+0.000$    & $+0.055$   &  $\bm{-0.285}$  & $-0.000$     & $\bm{+0.649}$ \\

\cline{4-13}
%----------------------------------------------------------------------------------------------------------------------------------------------
         &      &       &    $-x$    & $+0.851(8) $ & $-0.07(2)  $ & $-0.10(1)  $  & $+0.869$  &$-0.000$    & $+0.000$   &  $+0.816$  &$+0.000$    & $+0.000$ \\
         &      &       &    $-y$    & $-0.02(2)  $ & $+0.10(2)  $ & $+0.02(2)  $  & $+0.000$  &$+0.055$    & $+0.000$   &  $\bm{-0.285}$  & $\bm{+0.649}$   & $-0.000$ \\
         &      &       &    $-z$    & $-0.03(2)  $ & $+0.02(2)  $ & $-0.05(2)  $  & $+0.000$  &$+0.000$    & $-0.055$   &  $\bm{-0.285}$  & $-0.000$    & $\bm{-0.649}$ \\

\hline
%----------------------------------------------------------------------------------------------------------------------------------------------
  $+0.0$ &$+1.0$& $+0.5$&    $+x$    & $-0.876(3) $ & $+0.037(7) $ & $+0.084(7) $  & $-0.869$  &$+0.000$    & $+0.000$   &  $-0.846$  &$+0.000$    & $-0.000$ \\
         &      &       &    $$+y$$    & $-0.078(7) $ & $-0.485(6) $ & $-0.083(7) $  & $+0.000$  &$-0.544$    & $+0.000$   &  $\bm{+0.148}$  &$\bm{-0.798}$    & $-0.000$ \\
         &      &       &    $+z$    & $+0.069(7) $ & $-0.062(7) $ & $+0.481(6) $  & $+0.000$  &$+0.000$    & $+0.544$   &  $\bm{+0.148}$  & $+0.000$    & $\bm{+0.798}$ \\
                                         
\cline{4-13}
%----------------------------------------------------------------------------------------------------------------------------------------------
         &      &       &    $-x$    & $+0.880(3) $ & $-0.009(7) $ & $-0.044(7) $  & $+0.869$  &$-0.000$    & $-0.000$   &  $+0.886$  &$+0.000$    & $+0.000$ \\
         &      &       &    $-y$    & $+0.010(7) $ & $+0.495(6) $ & $+0.079(7) $  & $+0.000$  &$+0.544$    & $+0.000$   &  $\bm{+0.148}$  &$\bm{+0.798}$    & $+0.000$ \\
         &      &       &    $-z$    & $-0.101(7) $ & $+0.070(7) $ & $-0.478(6) $  & $+0.000$  &$+0.000$    & $-0.544$   &  $\bm{+0.148}$  & $+0.000$    & $\bm{-0.798}$ \\
                                                                                                              
\hline\hline                                                                                                        
\end{tabular}                                                                                                 
}                                                                                                             
\end{table*}

\subsection{Commensurate magnetic structure}\label{SNP_C}

The final polarization vector after the scattering process at the sample is described by the Blume-Maleyev-equations\cite{Blume:63,Maleyev:63} that may be given in the following convenient form
\begin{equation}\label{polarizationtensor.eqn}
\bm{P'}=\bm{\tilde{P}}\bm{P_0}+\bm{P''},
\end{equation}
where $\mathbf{\tilde{P}}$ is the polarization tensor, which describes the rotation of the initial polarization vector $\bm{P_0}$ in the scattering process and $\bm{P''}$ is the polarization created in the scattering process at the sample. For purely magnetic reflections, like the ones observed in NdFe$_3$($^{11}$BO$_3$)$_4$, $\bm{\tilde{P}}$ and $\bm{P''}$ are reduced to
\begin{align}\label{polarizationtensordetails.eqn}
\sigma\bm{\tilde{P}}&=\left(\begin{array}{ccc}
\scriptstyle-\vert\bm{M}_{\perp}\vert^2&\scriptstyle 0&\scriptstyle 0\\
\scriptstyle 0&\scriptstyle\vert M_{\perp y}\vert^2-\vert M_{\perp z}\vert^2&\scriptstyle 2\Re(M_{\perp y}^\ast\cdot M_{\perp z}) \\
\scriptstyle 0&\scriptstyle
2\Re(M_{\perp y}^\ast\cdot M_{\perp z})&\scriptstyle-\vert M_{\perp y}\vert^2+\vert M_{\perp z}\vert^2
\end{array}\right),
\\
\sigma\bm{P''}&=\left(\begin{array}{c}\scriptstyle-2\Im(M_{\perp y}^\ast\cdot M_{\perp z})\\\scriptstyle 0\\\scriptstyle0\end{array}\right),
\\
\sigma&=\vert\bm{M}_{\perp}\vert^2+P_{0x}2\Im(M_{\perp y}^\ast\cdot M_{\perp z}).\label{eqn:pol_cross}
\end{align}
Here $\bm{M}_{\perp}$ is the magnetic interaction vector defined as $\mathbf{M_{\perp}}=\hat{\mathbf{Q}}\times(\mathbf{\rho}(\mathbf{Q})\times\hat{\mathbf{Q}})$, where $\bf{\rho}(\mathbf{Q})=-2\mu_B\int\mathbf{\rho}(\mathbf{r})\exp(i\bf{Q}\cdot\mathbf{r})d\mathbf{r}$ is the Fourier transform of the magnetization density $\mathbf{\rho}(\bf{r})$ of the investigated sample and $\hat{\mathbf{Q}}$ is a unit vector parallel to the scattering vector $\mathbf{Q}$. The set of polarization axes is defined to have $x$ parallel to $\mathbf{Q}$, $z$ perpendicular to the scattering plane and $y$ completing the right-handed set. We note, that the term $2\Im(M_{\perp y}^\ast\cdot M_{\perp z})\equiv C$ is only non-zero for magnetic structures that display chirality and is therefore often denoted as the chiral term. Finally, the measured quantity is the polarization matrix, namely the components of the final polarization vector after the scattering process for all three directions of the incident beam  polarization,
\begin{equation}
\mathrm{P}_{ij}=(P_{i0}\tilde{P}_{ji}+P''_j),\label{eq:P_ij}
\end{equation}
where $i$ and $j$ ($i,j=x,y,z$) denote the directions of the incident and final polarization vectors, respectively.\\ 
For the commensurate magnetic phase of NdFe$_3$($^{11}$BO$_3$)$_4$ we measured the polarization matrix at six magnetic Bragg reflections and at T = 20, 25 and 30~K. The measured polarization matrices proved to be independent of temperature (within the error bars) for T~$\geq$~20~K. Thus, only the data for T~=~20~K will be discussed in the following. The corresponding matrices are provided in table \ref{tab:polartensors}.
From the measured polarization matrices several constraints on the magnetic structure can be derived:
\begin{enumerate}
\item On all measured Bragg peaks the elements $yx$ and $zx$ are equal to zero. This implies that the magnetic structure is not chiral at all or that it is a chiral structure with equally populated chiral domains.
\item For the magnetic reflection (0,0,-1.5) the scattering vector $\bm{Q}$ is directed parallel to the crystallographic c direction. Hence, the magnetic interaction vector only contains components in the basal plane. $\bm{P'}_{xx}=-\vert\bm{M}_{\perp}\vert^2/\vert\bm{M}_{\perp}\vert^2$~=~$-0.872(2)$. Here the reduction from $-1$ is due the polarizing benders which have a non-ideal polarization efficiency of approximately 0.966. Thus, $\bm{P'}_{xx}$ is fully polarized whereas $\bm{P'}_{yy} =(\vert M_{\perp y}\vert^2-\vert M_{\perp z}\vert^2)/\vert\bm{M}_{\perp}\vert^2$ and $\bm{P'}_{zz} (-\vert M_{\perp y}\vert^2+\vert M_{\perp z}\vert^2)/\vert\bm{M}_{\perp}\vert^2$ are fully depolarized. This would foremost lead to the assumption $\vert M_{\perp y}\vert^2\approx\vert M_{\perp z}\vert^2$ but as the elements $\bm{P'}_{yz}$ and $\bm{P'}_{zy}$ are also equal to zero it also suggests the presence of spin domains in the basal plane. 
\item On the magnetic reflection (0,4,0.5) the scattering vector $\bm{Q}$ is approximately parallel to the reciprocal b$^{\star}$ axis. As the z axis which is perpendicular to the scattering plane lies always within the basal hexagonal plane in the chosen scattering geometry the y-axis is approximately parallel to the crystallographic $c$-axis ($\sphericalangle(y,c)\approx15^{\circ}$). The polarization tensor shows $\bm{P'}_{xx}\approx\bm{P'}_{yy}\approx-\bm{P'}_{zz}$ which indicates that $\vert M_{\perp y}\vert^2\approx$~0 and hence the magnetic interaction vector is directed along z. Therefore the magnetic moments are confined in the basal plane.
\end{enumerate}
We note that these constraints are satisfied by both possible magnetic structures (M1a/b) and (M2). In order to calculate the expected polarization matrices for both models, we used lattice constants and structural parameters from table~\ref{tab:structure}. The magnitude of the magnetic moments for the Fe and Nd ions were set to the values as obtained from the unpolarized diffraction data. Since SNP is generally insensitive to absolute moment sizes in case of pure magnetic reflections they were fixed in subsequents fits. Both models fail to explain the observed polarization matrices when no orientation domains were considered. Introducing the three or six orientation domains with statistical population in the calculation for models (M1) and (M2), respectively (cf. appendix), resulted in polarization matrices given in table~\ref{tab:polartensors}. Further, we considered the non-ideal polarization efficiency of the used polarizing benders in the calculation.\\
As demonstrated in table~\ref{tab:polartensors} the magnetic model (M2) was not able to explain the measured polarization tensors on the two magnetic reflections  (0, 1, -2.5) and (0, 1, 0.5) as indicated by the bold entries in  table~~\ref{tab:polartensors}($\chi^2$~=~9.5). The model introduces small chiral contributions ($yx$ and $zx$ elements of the tensors) due to  its slightly canted spins (cf Fig.~\ref{Fig:old_mag_struc}(b)), that are not observed in the experiment. Performing a fit from this starting values did not result in a better agreement between model and data, as the fit diverged. Hence, model (M2) can be excluded.\\ 
For the magnetic model (M1) we fixed $\phi_{\rm{Fe}}$~=~0 and left $\phi_{\rm{Nd}}$ free for the fits, thus corresponding to (M1b). The fit converges to a solution with $\phi_{\rm{Nd}}$~=~0 ($\chi^2$~=~3.8) which corresponds to (M1a). Attempts to additionally determine the orientation of the Fe magnetic moments via the angle $\phi_{\rm{Fe}}$ gave no conclusive results. The angle $\phi_{\rm{Fe}}$ describes the absolute orientation of the magnetic moments in the crystallographic ab-plane with respect to chemical structure. Thus, the indeterminacy of $\phi_{\rm{Fe}}$ is most probably related to the presence of the three orientation domains in the hexagonal basal plane.
\\
In summary our SNP data for the commensurate phase is best explained by the model (M1a) when $\phi_{\rm{Fe}}$~=~$\phi_{\rm{Nd}}$~=~0 which is also in agreement with the unpolarized neutron single crystal diffraction results described in section~\ref{unpol_magstruc}. The corresponding magnetic structure is illustrated in Fig.~\ref{Fig:old_mag_struc}(a).\\
\begin{figure}[th!]
\includegraphics[width=0.35\textwidth]{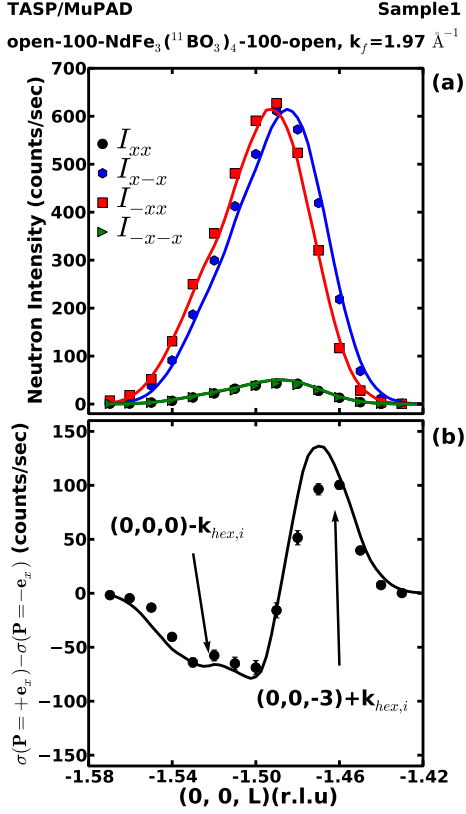}
\caption{\small{(a) $\bm{Q}$-scans in the polarized mode over the magnetic Bragg satellites (0,0,0)-$\bm{k}_{hex,i}$ and 0,0,-3)+$\bm{k}_{hex,i}$ at T~=~1.5~K are shown. (b) The difference for intensities with the initial polarization vector directed parallel and antiparallel to $\bm{Q}$ for the same $\bm{Q}$-scan are shown. This difference is directly proportional to the chiral contribution to the scattering cross-section as shown in Eq.~\eqref{eq:polx_chiral}. We note, that the installed supermirror benders simultaneously act as 100' collimators in front of and behind the sample, respectively.}}\label{fig:chiral}
\end{figure}
\begin{table}[th]                                                                                          
\caption{(a) Integrated intensities for the magnetic satellite reflections in the IC phase as calculated from model (M1a), however, with the incommensurate $\bm{k}_{hex,i}$. For the calculation we assumed that only one single chiral domain is populated. (b) Measured integrated intensities for the $\bm{Q}$-scan in Fig.~\ref{fig:chiral}(a) in the different polarization channels. For the comparison see the text.}
\label{tab:chiral_scan_ints}
\vspace{2mm}
\centering  
 {\footnotesize
 \begin{tabular}{p{4cm}|p{1cm}p{1cm}p{1cm}p{1cm}}                                                                          
 \multicolumn{5}{l}{(a)}\\ 
 \hline                                                                                                     
  Peak                   & $I^c_{xx}$  & $I^c_{x-x}$ & $I^c_{-xx}$ & $I^c_{-x-x}$\\\hline
%---------------------------------------------------------------------   
(0,0,0)-$\bm{k}_{hex,i}$ & 4.15 &       0.17 &      99.56 &      4.15 \\
(0,0,-3)+$\bm{k}_{hex,i}$& 4.16 &       99.83 &      0.17 &      4.16 \\
\hline

%---------------------------------------------
\end{tabular}    
\begin{tabular}{p{4cm}|p{1cm}p{1cm}p{1cm}p{1cm}}                                                                          
 \multicolumn{5}{l}{}\\ 
 \multicolumn{5}{l}{(b)}\\ 
 \hline                                                                                                     
 Peak                    & $I^m_{xx}$  & $I^m_{x-x}$ & $I^m_{-xx}$ & $I^m_{-x-x}$ \\\hline
 %---------------------------------------------------------------------   
 integrated over $\pm\bm{k}_{hex,i}$ & 66(8)      & 934(31)     & 946(31)     & 65(8) \\
 \hline
 %---------------------------------------------
 \end{tabular}
 }
\end{table}
\subsection{Chirality in the incommensurate magnetic phase}\label{SNP_IC}
Fig.~\ref{fig:chiral}(a) shows polarized constant-energy-scans along the L-direction performed around the reciprocal space position (0, 0, -1.5) at T~=~1.5~K in the IC-phase. Four different polarization channels were measured, namely $I_{xx}$,  $I_{x-x}$, $I_{-xx}$ and $I_{-x-x}$. The scans for the channels $I_{x-x}$ and $I_{-xx}$ appear to be slightly shifted towards the positions of the magnetic satellite reflections (0,0,-3)+$\bm{k}_{hex,i}$ ($I_{x-x}$) and (0,0,0)-$\bm{k}_{hex,i}$ ($I_{-xx}$), respectively, in agreement with the observations made in the high resolution measurement in section~\ref{highres}.\\
The effect is more significant in the sum of the four terms given by
\begin{eqnarray}
 \Delta I&=&(I_{xx} + I_{x-x})-(I_{-xx} + I_{-x-x})\nonumber\\
 &=&\sigma(P_{0x}=+1)-\sigma(P_{0x}=-1),\label{eqn:difference}
\end{eqnarray}
that is equivalent of measuring the difference of the two polarized neutron cross-sections with the initial polarization vector $\bm{P}_0$ directed parallel and antiparallel to the $x$ direction (thus parallel or antiparallel to $\bm{Q}$). The resulting scan is shown in Fig.~\ref{fig:chiral}(b).\\
By means of Eqns.~\eqref{eqn:pol_cross} and~\eqref{eqn:difference} we further obtain the result
\begin{eqnarray}
 \Delta I&=&(\vert\bm{M}_{\perp}\vert^2+C)-(\vert\bm{M}_{\perp}\vert^2-C)\nonumber\\
 &=&2C,\label{eq:polx_chiral}
\end{eqnarray}
demonstrating that $\Delta I$ is proportional to the chiral term. Thus, by carrying out the polarized scans we were able to separate the resolution limited peak at (0,0,-1.5) in two satellites (0,0,0)-$\bm{k}_{hex,i}$ (I$_{-xx}$) and (0,0,-3)+$\bm{k}_{hex,i}$ (I$_{x-x}$). Since two distinct extrema with opposite signs for the two satellites can be distinguished the collinear antiferromagnetic structure of NdFe$_3$($^{11}$BO$_3$)$_4$ observed in the commensurate phase appears to transform to an antiferromagnetic long-period spiral below T$_{IC}$. In addition the non-zero chiral term also suggests that at least unequally populated chiral domains are present in NdFe$_3$($^{11}$BO$_3$)$_4$.\\
In order to investigate this in more detail we performed calculations based on the magnetic model (M1a), however together with the incommensurate propagation vector $\bm{k}_{hex,i}$ that leads to a small rotation of the magnetic moments between neighboring layers. We calculated the integrated intensities for the four measured polarization channels $I_{xx}$,  $I_{x-x}$, $I_{-xx}$ and $I_{-x-x}$ for both satellites (0,0,0)-$\bm{k}_{hex,i}$ and (0,0,-3)+$\bm{k}_{hex,i}$, where we used $\varepsilon$ = 0.0667 in $\bm{k}_{hex,i}$ = [0, 0, $\frac{3}{2}$ + $\varepsilon$] as deduced from our high resolution diffraction data. The calculated integrated intensities are given in table \ref{tab:chiral_scan_ints}(a) for the case of only a single chirality domain being populated. In addition we assumed only one orientation domain, similar as for the unpolarized results. For comparison we give the measured integrated intensities from the scans for the four polarization channels in table \ref{tab:chiral_scan_ints}(b). Due to the limited resolution of the setup the measured values for the individual polarization channels are integrated over both peaks and cannot be compared directly to the calculations. This is especially true for the difference of the intensities. However, two reasonable assumptions can be made to allow for a comparison:
\renewcommand{\labelenumi}{(\roman{enumi})}
\begin{enumerate}
 \item the measured integrated intensity in the channel `$-xx$' is only due to the peak at (0,0,0)-$\bm{k}_{hex,i}$ whereas the channel `$x-x$' is only due to the peak at (0,0,-3)+$\bm{k}_{hex,i}$. This assumption is justified as the measured points of each respective polarization channel (red squares and blue circles in Fig.~\ref{fig:chiral}(a)) are shifted towards the direction of the corresponding peak.
 \item the channels `$xx$' and `$-x-x$' are equally contributed by both satellite reflections as they are centered on top of each other and between the other two channels (black circles and green triangles in Fig.~\ref{fig:chiral}(a)).
\end{enumerate}
Based on these assumptions we are able to calculate the ratio $I^m_{x-x}/(I^m_{xx}/2)$ = 28(4) for the intensities measured on the magnetic Bragg reflection (0,0,-3)+$\bm{k}_{hex,i}$. Note that the division by 2 is due to assumption (ii). The same ratio for the calculated intensities for this peak amounts to $I^c_{x-x}/I^c_{xx}$~=~24. For the satellite (0,0,0)-$\bm{k}_{hex,i}$ we obtain the measured ratio $I^m_{-xx}/(I^m_{-x-x}/2)$ = 28(4) and the calculated ratio $I^c_{-xx}/I^c_{-x-x}$ = 24. This indicates that our model of a long-period antiferromagnetic helix propagating along the hexagonal $c$-axis with the magnetic moments parallel to the hexagonal basal plane and single chirality domain is in good agreement with our data. \\ 
In addition we verified our assumptions by performing a convolution of the calculated integrated intensities for each of the four polarization channels (cf. table \ref{tab:chiral_scan_ints}) with the four-dimensional resolution function of the spectrometer in a simulation. The results of the simulations are the solid lines in Fig.~\ref{fig:chiral}(a). To match the intensity a single scale factor four all four polarization channels was introduced in the calculation. Similar as for the previous calculations the simulation was performed with only one of the two chiral domains being populated, and a single orientation domain. The solid line in Fig.~\ref{fig:chiral}(b) was obtained by summating the individual curves for each polarization channel with respect to Eq.~\eqref{eqn:difference}. The peculiar shape of difference curve is due to the slightly asymmetric shape of the magnetic satellite peaks. After the asymmetric peak shape was taken account for within the simulation good agreement between simulation and experimental data was achieved.

\section{Discussion}
The present single-crystal neutron diffraction investigations did not detect significant deviations from space group \textit{R32} concerning the chemical structure of multiferroic NdFe$_3$($^{11}$BO$_3$)$_4$ at low temperatures. With respect to magnetic ordering this study shows that only the magnetic model (M1a) is in agreement with our polarized neutron data on the noncentrosymmetric NdFe$_3$($^{11}$BO$_3$)$_4$.\\ 
However, in contrast to the previous neutron powder diffraction study of NdFe$_3$($^{11}$BO$_3$)$_4$\cite{fischer:06}, we may conclude from our combined investigation with unpolarized and polarized neutrons that in addition, ferromagnetic alignment of the magnetic Fe and Nd sublattices holds. This implies low magnetic symmetry such as \textit{R12} (monoclinic \textit{C2}) in the commensurate phase. Our conclusion is further supported by measurements of the magnetic susceptibility by Tristan et al.\cite{Tristan:07} that yield an easy magnetization along the $a$-axis. In addition, we should emphasize that not only Fe$^{3+}$, but also the Nd$^{3+}$ ions show antiferromagnetic long-range order below the N\'{e}el temperature in case of neodymium ferroborate.\\
Concerning the incommensurate phase the polarized $\bm{Q}$-scans over the positions of the magnetic peaks could be well explained via the magnetic model (M1a) that was found for the commensurate phase simply by introducing the incommensurate propagation vector $\bm{k}_{hex,i}$. This suggests that the magnetic structure transforms into a long-period antiferromagnetic helix that propagates along the hexagonal $c$-axis with the magnetic moments perpendicular to it. The incommensurate magnetic propagation vector $\bm{k}_{hex,i}$ = [0, 0, $\frac{3}{2}$ + $\varepsilon$] is therefore associated with a rotation of the magnetic moments about 180$^\circ$ + $\gamma$ around the $c$-axis between adjacent hexagonal planes that are interrelated via trigonal translations. The measured value of the splitting $\varepsilon$ = 0.00667 corresponds to $\gamma$~$\approx$~0.8$^\circ$ and the full period of the helix amounts to approximately 1140~\AA. The mere observation of a chiral contribution by means of polarization analysis signifies unequally populated chirality domains. Our data further suggest that only one of the two chirality domains is populated. The antiferromagnetic helix in the IC phase of NdFe$_3$($^{11}$BO$_3$)$_4$ therefore exists with an unique handedness. A single chirality domain is in principle not expected, since left- and right-handed spirals are energetically degenerate, however in the case of NdFe$_3$($^{11}$BO$_3$)$_4$ this might be related to the fact that the chemical structure is non-centrosymmetric. This is similar to the magnetic spirals in MnSi or UPtGe\cite{brown:01}, and the more recent example Ba$_3$NbFe$_3$Si$_2$O$_{14}$\cite{Marty:08} that all three possess no inversion symmetry. In particular the Jana2006 analysis has shown that in addition the incommensurate magnetic structure exists in a single magnetic orientation domain with full superspace group symmetry \textit{R32(00$\gamma$)t0} and in full agreement with the polarized results.\\
The determined saturation value of the magnetic Fe moment of approximately 4.2~$\mu_B$ (see Fig.~\ref{fig:magmom}) is less than 5~$\mu_B$ which would be expected for a free Fe$^3+$ ion, in contrast to our previous powder results\cite{fischer:06}. Moreover, the ferromagnetic alignment of the Fe and Nd magnetic moments implies considerably smaller magnitudes for Nd than derived from the powder diffraction data. The latter are not affected by extinction. The difference may be to a certain extent due to the extinction effects, as the even smaller saturation value of Fe in case of thermal neutrons is caused by the considerably larger extinction, compared to hot neutrons. On the other hand, for TbFe$_3$($^{11}$BO$_3$)$_4$ at 2 K,   Ritter at al. determined by means of powder neutron diffraction an antiparallel alignment of the Fe and Tb magnetic moments with ordered magnitudes $\mu_{\rm{Fe}}$ = 4.39(4) $\mu_B$ and $\mu_{\rm{Tb}}$ = 8.53(5) $\mu_B$\cite{ritter:07}. Despite the bondvalence result $3+$ for Fe derived by these authors, the ordered Fe magnetic moment is also reduced. Hence, another possible reason could be frustration effects, yielding partially disordered magnetic moments.\\
Furthermore, the observation of third order harmonics of the magnetic satellites at the positions (0, 0, 3/2 $\pm$ 3$\varepsilon$) in the incommensurate phase additionally suggest the formation of a magnetic soliton lattice in NdFe$_3$($^{11}$BO$_3$)$_4$\cite{roessli:01}. A soliton is the appearance of localized or topological defects in periodic structures due to the presence of non-linear forces. Such non-linear forces can be due to an external magnetic field that interacts with the magnetic moments or due to magnetic anisotropy as shown theoretically by Izyumov and Laptev\cite{Izyumov:83}. Indeed NdFe$_3$($^{11}$BO$_3$)$_4$ exhibits magnetic anisotropy in the hexagonal basal plane as demonstrated by the results of Tristan et al.\cite{Tristan:07}. The calculation of Izyumov and Laptev is based on a magnetic helix that forms due to the the presence of the Dzyaloshinsky-Moriya interaction (DMI). Until now no explicit statement about the existence of the DMI in NdFe$_3$($^{11}$BO$_3$)$_4$ has been made. However, as NdFe$_3$($^{11}$BO$_3$)$_4$ is non-centrosymmetric the presence of the DMI is allowed from symmetry. Therefore, the formation of the observed magnetic helix is possibly driven by the DMI.\\
In Figs.~\ref{Fig:nd_second_order_sats} (b) to (d) we see that the intensities of the second order satellites are highest for temperatures T~$\lesssim$~13.5~K and the distortions of the incommensurate periodic structures seem to be largest near to the C-IC phase transition. We therefore assume that at T$_{IC}$ the interaction that favors a magnetic order that is incommensurate with respect to the underlying crystal lattice becomes non-negligible and leads to non-linear forces onto the magnetic subsystem slightly below T$_{IC}$ as it still wants to remain in its commensurate magnetic order. Within a small temperature regime below T$_{IC}$ the magnetic structure consequently is not yet completely incommensurate but can be rather viewed as a distorted commensurate magnetic structure with domain walls. Alternating periods of commensurate parts and domains walls then lead to the observed third order harmonics. The observation of a magnetic soliton lattice without the application of external forces like magnetic fields or mechanical stress are rather unlikely and to the best of our knowledge the only other compound for which a magnetic soliton lattice was reported without the application of an external magnetic field is CuB$_2$O$_4$~\cite{roessli:01}. The observed temperature dependence of the propagation vector is continuous and described by Eq.~\eqref{eq:k_temp_dep}. This behavior is  close to $k(T) \propto \vert(T_{IC}-T)\vert^{0.48}$ reported in Ref.~\onlinecite{roessli:01}. In addition, similar to CuB$_2$O$_4$ the commensurate phase is realized when the temperature is increased, which is in contradiction to the prediction of the theory \cite{Izyumov:83}. For CuB$_2$O$_4$ it was proposed that the difference to the  theory can be explained by assuming that the change of the propagation vector is not due to a temperature dependent magnetic anisotropy as in Ref.~\onlinecite{Izyumov:83} but rather due to the magnitude of the DMI that decreases as a function of increasing temperature~\cite{boehm:02}. We assume that is is similarly true for NdFe$_3$($^{11}$BO$_3$)$_4$. In summary our experimental results are well described by the assumption of a magnetic soliton lattice.\\
The origin of the less pronounced incommensurability in case of the small crystal studied (Sample2) in the present work, compared to the larger one (Sample1) that was also used in Ref.~\onlinecite{fischer:06}, is not yet clear and should be clarified by future systematic investigations of possible sample dependencies.
\section{Conclusion}
Our neutron diffraction results show that the long-range magnetic order of multiferroic NdFe$_3$($^{11}$BO$_3$)$_4$ observed below $T_N\approx$~30~K consists of antiferromagnetic stacking along the $c$-axis, where the magnetic moments of all three Fe$^{3+}$ sublattices and the Nd$^{3+}$ sublattice are aligned ferromagnetically and parallel to the hexagonal basal plane, corresponding to model (M1a). Below T$_{IC}\approx$~13.5~K the magnetic structure turns into an incommensurate antiferromagnetic helix propagating along the $c$-axis with a period of approximately 1140~\AA.\\
Our polarized neutron diffraction data further suggests that the helix is monochiral, i.e. only one of the two possible chiral domains is fully populated. The single magnetic chirality in neodymium ferroborate can be explained in terms of its non-centrosymmetric chemical structure, similar as for MnSi\cite{brown:01}, and the more recent example Ba$_3$NbFe$_3$Si$_2$O$_{14}$\cite{Marty:08}. To the best of our knowledge the former two materials are the only examples apart from NdFe$_3$($^{11}$BO$_3$)$_4$ that show this peculiar property.\\
In the case of NdFe$_3$($^{11}$BO$_3$)$_4$ the commensurate-incommensurate magnetic phase transition is possibly accompanied by the formation of a magnetic soliton lattice, as indicated by the observation of third order harmonics of the magnetic Bragg peaks. This further suggests that the magnetic helix in NdFe$_3$($^{11}$BO$_3$)$_4$ may be driven by the Dzyaloshinskii-Moriya interaction, which would be allowed by symmetry.\\
In conclusion, we identified the new monochiral compound NdFe$_3$($^{11}$BO$_3$)$_4$ that provides us with a new model system to investigate the interesting properties of magnetic chirality in condensed matter.

\begin{table*}[th!]                                                                                          
\begin{center}
\caption{Magnetic refinements of NdFe$_3$($^{11}$BO$_3$)$_4$ by means of FullProf\cite{fullprof}. Here the three character encoding (column set in the table) of the different fit runs is as follows. The first letter 'F' signifies that the fit was performed with FullProf (cf. Jana2006 refinements in table~\ref{tab:jana_ref}), the number at second position indicates which data set has been used, the characters at the third position characterize constraints that were applied. Here a and b indicate that the polar angle $\phi_{\rm{Nd}}$ was fixed to zero and left free for the fits, respectively (cf. cases M1a and M1b in section~\ref{unpol_magstruc}).}
\label{tab:fullprof_ref}
 {\footnotesize
 \begin{tabular}{cccc|c|cccc|ccc|c}\hline                                                                          
 Instrument & T (K) & Phase & \# refl.  & Set & R$_{m,F2}$ (\%)   & R$_{m,F2w}$ (\%)& R$_{m,F}$(\%) & $\chi^2$ & $\mu$(Fe) ($\mu_B$) & $\mu$(Nd) ($\mu_B$) & $\phi_{\rm{Nd}}$ ($^{\circ}$) & \# domains  \\\hline\hline
%---------------------------------------------------------------------------------------------------
 HEiDi      & 22.5  & C     & 82       & F1a & 15.9             & 9.2             & 14.8          & 1.07     & 3.180(8)            & 0.32(2)             &0     & 3  \\\hline 

%---------------------------------------------------------------------------------------------------
	    &       &         &        & F1b & 16.0             & 9.1             & 14.1          & 1.05     & 3.132(6)            & 1.03(5)             & 68.5(8) & 6  \\\hline\hline
%---------------------------------------------------------------------------------------------------
%---------------------------------------------------------------------------------------------------
 HEiDi      & 15   & C     & 50        & F2a  & 8.9              & 6.6             & 5.9           & 2.88     & 3.910(9)   & 0.55(2) & 0    & 3 \\\hline
%---------------------------------------------------------------------------------------------------
	    &       &         &        & F2a'  & 8.9              & 6.6             & 5.9           & 2.89    & 3.911(4) & 0.58(1) & 15$^1$ & 3 \\\hline
%---------------------------------------------------------------------------------------------------
	    &       &         &        & F2a''  & 8.9              & 6.6             & 6.0           & 2.90 & 3.881(6) & 1.14(3) & 30$^1$ & 3 \\\hline\hline
%---------------------------------------------------------------------------------------------------
%---------------------------------------------------------------------------------------------------
 HEiDi      &  5   & IC     & 50        & F3a  & 6.9              & 5.1             & 4.6           &               	   2.62 & 4.22(3) & 1.13(6) & 0     & 1 \\\hline
%---------------------------------------------------------------------------------------------------
	    &       &         &        & F3b  & 6.8              & 5.1             & 4.6         & 2.68 & 4.22(2) & 1.3(2) & 28(15) & 2 \\\hline\hline
%---------------------------------------------------------------------------------------------------
%---------------------------------------------------------------------------------------------------
 TriCS      &  5.5   & IC     & 152    & F4a  & 8.7              &  9.6            & 9.9         &          	   43 & 3.73(2) & 0.93(3) & 0     & 1 \\\hline
%---------------------------------------------------------------------------------------------------
	    &       &         &        & F4b  & 7.2              &  8.2            & 6.6 	 &	31 & 3.659(9) & 1.34(2) & 44.5(9) & 2 \\\hline\hline
%---------------------------------------------------------------------------------------------------
%---------------------------------------------------------------------------------------------------
\end{tabular}}
\end{center}
\begin{flushleft}
$^1$ $\phi_{\rm{Nd}}$ was fixed at the given value different from $\phi_{\rm{Nd}}$~=~0.
\end{flushleft}

\end{table*}
\begin{acknowledgments}
Partially this work has been performed at the Swiss spallation neutron source SINQ\cite{fischer:97} (instruments TriCS\cite{schefer:00}, TASP\cite{semadeni:01}, MuPAD\cite{janoschek:07}) and another part at the single crystal neutron diffractometer HEiDi\cite{meven:05} for hot neutrons, situated at the Forschungsneutronenquelle Heinz Maier-Leibnitz (FRM II). Further we are thankful to Severian Gvasaliya for providing experimental support for the measurements performed on TASP. MJ is grateful to Ben Taylor and Dominik Bauer for useful discussions.
\end{acknowledgments}

\appendix

\section{Characteristic results from FullProf refinements}
Here we want to discuss the results of our magnetic refinements for NdFe$_3$($^{11}$BO$_3$)$_4$ by means of FullProf in more detail. We have chosen several characteristic data sets that highlight the results of the fits provided in table~\ref{tab:fullprof_ref}.\\
First we will discuss the data sets measured on HEiDi with hot neutrons ($\lambda$~=~0.55 \AA). For the commensurate magnetic phase of NdFe$_3$($^{11}$BO$_3$)$_4$ we performed fits with $\phi_{\rm{Nd}}$ fixed to zero and free, respectively. The corresponding sets in table ~\ref{tab:fullprof_ref} are (F1a) and 
(F1b) at T~=~22.5 K, respectively. The refinements were carried out with the nuclear scale factor and extinction parameter 2.9(1) that was determined by the fit (N1) (cf. table~\ref{tab:structure}). The determined angle $\phi_{\rm{Nd}}$ for (F1b) is of similar magnitude as the value 76(3)$^{\circ}$ of the previous powder neutron diffraction results\cite{fischer:06}. However, according to Hamilton's signficance test\cite{hamilton:65}, one should consider the ratio 1.01 of the weighted R-factors. Then R$_{1,79,0.25}$~$\approx$1.01 from table 1 in Ref.~\onlinecite{hamilton:65} indicates only a marginal significance level of 0.25 that (F1b) is the correct result. Further refinements at lower temperatures support this as will be demonstrated in the following. Further, variation of also the extinction parameter yields the value 3.0(8) and $\chi^2$ = 1.07. Therefore in case of NdFe$_3$($^{11}$BO$_3$)$_4$ the nuclear and magnetic extinction parameters agree within error limits. Thus we use the former also for the refinements of the subsequent 'small' HEiDi data sets with 5 nuclear and 50 magnetic peaks. The former were used to obtain the scale factor.\\
At T~=~ 15 K we performed three different sets of fits where the angle $\phi_{\rm{Nd}}$ was consequently fixed to 0, 15 and 30 $^{\circ}$ (cf. sets (F2a,F2a',F2a'') in table ~\ref{tab:fullprof_ref}). There is apparently a very flat minimum centered at $\phi_{\rm{Nd}}$ = 0 which proves the assumption $\phi_{\rm{Nd}}$ = 0 to be correct.\\
The same is valid for the incommensurate magnetic phase of NdFe$_3$($^{11}$BO$_3$)$_4$. Fits of the data at 5 K show better agreement factor when $\phi_{\rm{Nd}}$ is fixed to zero (cf. sets (F3a+b) in table ~\ref{tab:fullprof_ref}).\\
Finally, the results obtained by fits of the TriCS data measured with thermal neutrons ($\lambda$~=~1.18\AA) at T~=~5.5~K gives the following results. The nuclear fit (N2) yielded the scale factor 40(1) and a considerably larger extinction parameter 18(1), compared to the hot neutron HEiDi
data. Best fits of the magnetic neutron intensities had been obtained with temperature parameter B = 0 of the metal atoms. The positional x-parameter of Fe had been taken from (N2), table ~\ref{tab:structure}. Good fits were only obtained by refining also the extinction parameter, resulting in the considerably smaller magnetic value 2.3(2), compared to the almost an
order larger nuclear value. The results of the refinement of the magnetic intensities is given in table ~\ref{tab:fullprof_ref} sets (F4a) and (F4b), corresponding again to fixed and free polar angle $\phi_{\rm{Nd}}$, respectively. For (F4b) the extinction parameter was fixed to the value from (F4a). The
latter does not change essentially, if it is varied too. Although refinement (F4a) seems with an additional parameter somewhat better, we think that it is hardly significant with respect to the comparable hot neutron 5 K results, which in principle are considerably less affected by extinction.
\begin{table*}[th]                                                                                          
\caption{Characteristic magnetic refinements by means of Jana2006\cite{jana2006}. Here the three character encoding (column set in the table) of the different fit runs is as for table~\ref{tab:fullprof_ref}). Thus, the first character 'J' indicates that the fits were performed with Jana2006. E.g. (J3x) means that the data is identical to the data set for the case (F3x) shown in table~\ref{tab:fullprof_ref}.}
\label{tab:jana_ref}
\centering                                                                                                
 {\footnotesize
 \begin{tabular}{cc|c|ccccc|ccc|c}\hline                                                                          
 Instrument & T (K) & Set & superspace group & R$_m$(all) & R$_{mw}$(all) & GOF(all)& GOF(obs)  & $\mu$(Fe) ($\mu_B$) & $\mu$(Nd) ($\mu_B$) & $\phi_{\rm{Nd}}$ ($^{\circ}$) & domains \\\hline\hline
%---------------------------------------------------------------------------------------------------
%---------------------------------------------------------------------------------------------------
 HEiDi      &  22.5    & J1a  & \textit{R12($\alpha$,2$\alpha$,$\gamma$)00} & 15.1 & 12.6 & 1.05 & 1.24 & 3.126(2) &  0.263(4) & 0 & 3 \\
	    &        &      &  ($\alpha$=0;$\gamma$=3/2)&&&&&&&&\\\hline
%---------------------------------------------------------------------------------------------------
	    &        & J1b  &\textit{R1($\alpha$,$\beta$,$\gamma$)0} & 14.1 & 12.1 & 1.02 & 1.13 & 3.044(3) &  1.29(2) & 73.7(2) & 6\\
	    &        &      &  ($\alpha$=$\beta$=0;$\gamma$=3/2)&&&&&&&&\\\hline\hline
%---------------------------------------------------------------------------------------------------
%---------------------------------------------------------------------------------------------------
 HEiDi      &  15    & J2a  & \textit{R12($\alpha$,2$\alpha$,$\gamma$)00} & 6.0 & 10.0 & 1.60 & 1.63 & 3.840(2) &  0.474(4) & 0 & 3 \\
	    &        &      &  ($\alpha$=0;$\gamma$=3/2)&&&&&&&&\\\hline
%---------------------------------------------------------------------------------------------------
	    &        & J2b  &\textit{R1($\alpha$,$\beta$,$\gamma$)0} & 6.3 & 10.0 & 1.62 & 1.65 & 3.782(3) &  1.27(3) & 65.2(5) & 6\\
	    &        &      &  ($\alpha$=$\beta$=0;$\gamma$=3/2)&&&&&&&&\\\hline\hline
%---------------------------------------------------------------------------------------------------
%--------------------------------------------------------------------------------------------------- 
 HEiDi      &  5    & J3a  & \textit{R32(00$\gamma$)t0} & 4.6 & 8.2 & 1.47 & 1.50 & 4.17(4) &  1.08(9) & 0 & 1 \\\hline
%---------------------------------------------------------------------------------------------------
	    &       & J3b  &\textit{R3(00$\gamma$)t} & 4.4 & 7.9 & 1.43 & 1.46 & 4.07(1) &  1.87(3) & 51.6(7) & 2\\\hline\hline
%---------------------------------------------------------------------------------------------------
%---------------------------------------------------------------------------------------------------
%--------------------------------------------------------------------------------------------------- 
 TriCS      &  5.5    & J4a  & \textit{R32(00$\gamma$)t0} & 10.6 & 19.5 & 5.60 & 5.60 & 4.03(5) &  1.15(6) & 0 & 1 \\\hline
%---------------------------------------------------------------------------------------------------
	    &         & J4b  &\textit{R3(00$\gamma$)t} & 7.3 & 15.6 & 4.48 & 4.48 & 3.832(2)&  1.536(7) & 44.9(2) & 2\\\hline\hline
\end{tabular}
}
\end{table*}
\section{Characteristic results from Jana2006 refinements}
In the following we will discuss our additional refinements that were performed with Jana2006\cite{jana2006}. Recently the option to refine magnetic structures has been implemented. It should be emphasized that magnetic structures have to be defined in a different way in Jana2006 compared to FullProf. In Jana2006 commensurate and incommensurate magnetic structures are described with superspace groups\cite{Janssen:06} similar to occupationally modulated chemical structures. The modulation function of the magnetic axial vector configuration is represented by means of a Fourier expansion according to the observed k-vectors, similar to Ref.~\onlinecite{lyons:62}. Moreover, the irreducible representations of the chemical structure with their little group $G_k$ and the associated magnetic basis functions as well as time inversion are taken into account. The atomic magnetic moments are described in a polar coordinate system.\\
The incommensurate and commensurate magnetic structures of NdFe$_3$($^{11}$BO$_3$)$_4$ correspond to a complex one-dimensional and to a two-dimensional representation, respectively. Keeping the chemical structure according to space group \textit{R32}, the former is described in by means of a four-dimensional superspace approach with cosine and sine waves associated with the k-vectors $\bm{k}_{hex,i}$ = [0,0,$\pm$1.502]. For ferromagnetic coupling of the magnetic Fe and Nd moments in the 'easy' (a,b)-plane of NdFe$_3$($^{11}$BO$_3$)$_4$, the magnetic superspace group \textit{R32(00$\gamma$)t0} was used. With the Nd moments deviating from $\phi_{\rm{Fe}}$ = 0 in the (a,b)-plane, magnetic superspace group \textit{R3(00$\gamma$)t} holds. Here the symmetry implies one or two magnetic domains, respectively. In the latter case they were found to be statistically populated.\\
In the magnetic commensurate case with $\bm{k}_{hex}$ = [0,0,3/2], being equivalent to the two-fold superstructure, only a cosine wave with magnetic superspace group \textit{R12($\alpha$,$\alpha$,$\gamma$)00} ($\alpha$=$\beta$=0;$\gamma$=3/2) is appropriate for ferromagnetic alignment of the magnetic Nd and Fe moments in the (a,b)-plane. For magnetic monoclinic superspace group \textit{R1($\alpha$,$\beta$,$\gamma$)t} ($\alpha$=$\beta$=0;$\gamma$=3/2) the magnetic Nd and Fe moments are not oriented parallel in the (a,b)-plane. The two cases imply three or six magnetic domains, respectively. They were found to be statistically populated.\\
As an example we discussed the results of combined refinement of 5 nuclear (to obtain the scale factor) and 50 magnetic peaks at 15 K (HEiDi measurements) that are given in table~\ref{tab:jana_ref}. Here we used the extinction parameter RhiIso~=~0.109(8) and the atom positions from table~\ref{tab:structure}. (J2a) and (J2b) correspond to fixed and free polar angle $\phi_{\rm{Nd}}$. The resulting R-factors of the refinement for this temperature show obviously no significance for an additional parameter $\phi_{\rm{Nd}}$ and again indicate that the ferromagnetic alignment of the Fe and Nd magnetic moments is correct.\\
For the data at 5 K a non-zero angle $\phi_{\rm{Nd}}$  shows a small improvement of the refinement (J3b), however, in view of the additional refinement parameter $\phi_{\rm{Nd}}$, one may question whether the small improvement is significant. Based on Hamilton's significance test\cite{hamilton:65}, the $R_{mw}$-ratio 8.2/7.9 = 1.04 should be considered. In cases (J3a) and (J3b)  we have 2 and three parameters, respectively and 55 reflections, i.e $R_{1,52,0.05}$~$\approx$~1.04 from table 1 in Ref.~\onlinecite{hamilton:65} indicates the correctness of $\phi_{\rm{Nd}}$=0 (J3a) at a significance level of approximately 0.05.\\
On the other hand (J4b) appears to be better than (J4a). Presumably this is related to the considerably larger extinction effects associated with thermal neutrons compared to hot neutrons. Apart from this temperature, the Jana2006 data evaluation yields values for the magnetic moments that agree within the error bars with the values shown in Fig.~\ref{fig:magmom} which is based on the FullProf evaluation. It should be noted that the superspace group description yields a natural explanation for the number of magnetic domains\cite{petricek:09}. The latter were found to be statistically occupied.
%
%\bibliography{ndfe_snp_arxiv.bib}

\end{document}